\documentclass[a4paper,11pt]{article}
\pdfoutput=1 % if your are submitting a pdflatex (i.e. if you have
\usepackage{jheppub} % for details on the use of the package, please
\usepackage[T1]{fontenc} % if needed
\RequirePackage[displaymath]{lineno} % Display line numbers

\usepackage{epsfig}
\usepackage{graphicx}% Include figure files
\usepackage{dcolumn}% Align table columns on decimal point
\usepackage{bm}% bold math
\usepackage{ltablex,booktabs}
\usepackage{overpic}
\usepackage{subfigure}
\usepackage{float}
\usepackage{color}
\usepackage{amsmath}
\usepackage{mathcomp}
\usepackage{mathrsfs}
\usepackage{multirow}
\usepackage{rotating}
\usepackage{amssymb}
\usepackage{gensymb}
\usepackage{amsmath}
\usepackage{tabularx}
\usepackage{epstopdf}

\usepackage{graphicx}
\usepackage{float}
\usepackage{units}
\usepackage{epsfig}
\usepackage{bigstrut}
\usepackage{overpic}
\usepackage{dcolumn}
\usepackage{bm}
\usepackage{color}
\usepackage{amsmath}
\usepackage{amsfonts}
\usepackage{lineno}
\usepackage{xspace}
\usepackage{multirow}
\usepackage{subfigure}
\usepackage{verbatim}
\usepackage{footmisc}
\usepackage{epstopdf}
\usepackage{tasks}
\usepackage{tasks}
\usepackage{xcolor}
\usepackage{soul}

\newcommand{\mev}{\,\unit{MeV}}

\newcommand{\mevcc}{\,\unit{MeV}/c^2}

\newcommand{\ee}{e^{+}e^{-}}

\newcommand{\pip}{\pi^+}
\newcommand{\pim}{\pi^-}

\newcommand{\Modec}{\overline{p}{}K_{S}^0\pi^0}

\newcommand{\Modedd}{\overline{\Lambda}{}\pi^-\pi^+\pi^-}

\newcommand{\Modeccc}{\overline{\Sigma}{}^{-}\pi^0}
\newcommand{\Modeddd}{\overline{\Sigma}{}^{-}\pi^-\pi^+}

\newcommand{\lambdacp}{\Lambda_{c}^{+}}
\newcommand{\lambdacm}{\overline{\Lambda}{}_{c}^{-}}

\newcommand{\LctoKsX}{\Lambda_{c}^{+} \to K_{S}^{0} X}

\newcommand{\LctoKzX}{\Lambda_{c}^{+} \to \overline{K}^{0} / K^{0} X}

\title{\boldmath Improved measurement of absolute branching fraction of the inclusive decay  $\LctoKsX$}
\collaborationImg{\includegraphics[width=0.15\textwidth, angle=90]{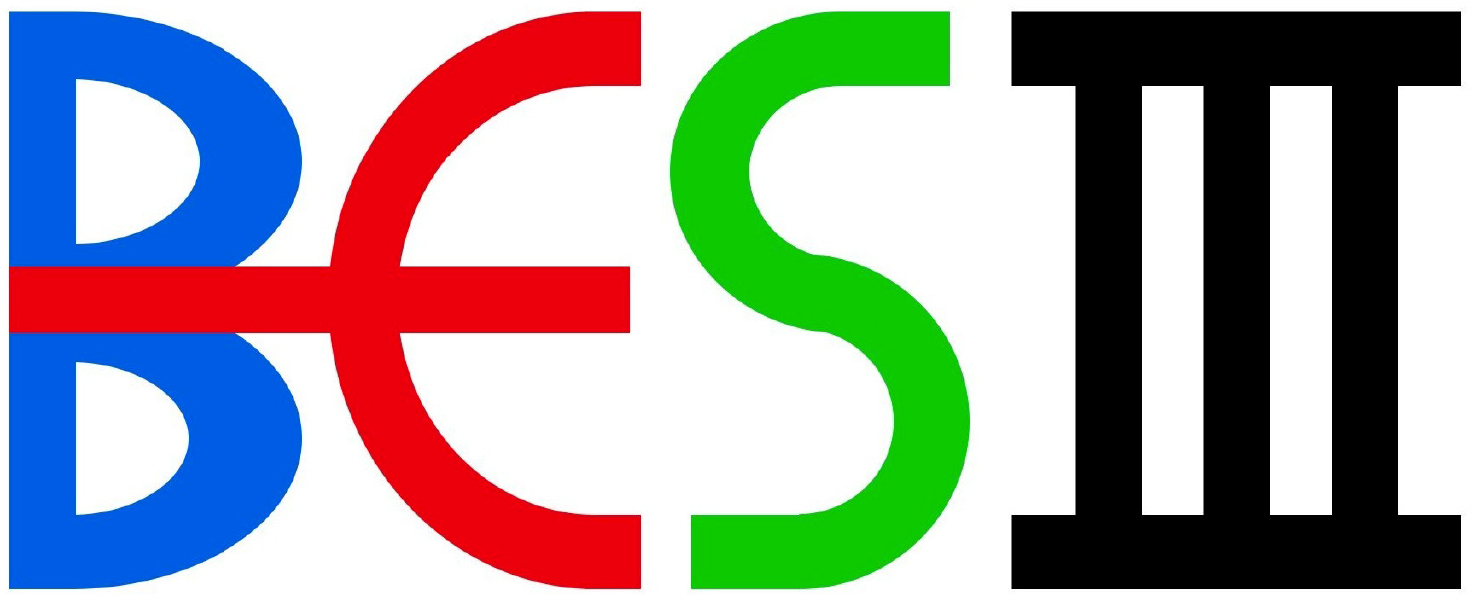}}
\collaboration{The BESIII Collaboration}

\date{\today} \abstract{By analyzing 4.5 fb$^{-1}$ of $\ee$
  collision data accumulated with the BESIII detector at
  center-of-mass energies ranging from 4599.53 MeV to 4698.82 MeV, we
  report the measurement of the absolute branching fraction (BF) of the
  inclusive decay $\LctoKsX$ using the double-tag technique. The
  result is $\mathcal{B}(\LctoKsX)=(10.9\pm0.2\pm0.1)\%$, where the
  first uncertainty is statistical and the second is systematic. This
  result indicates that there are still undiscovered decay channels
  containing $K_{S}^{0}$ in the final state with a combined BF of
  $(3.0\pm0.4)\%$. The BF of the inclusive decay $\LctoKzX$ is
  calculated to be $\mathcal{B}(\LctoKzX)=(21.8 \pm0.4 \pm0.2)\%$, 
  The result is
  in agreement with the prediction of the statistical isospin model.}
	
\arxivnumber{}
\begin{document}
%\linenumbers
\maketitle
\flushbottom

%------------------------------------------------------------------------------
\section{Introduction}
Charmed baryons serve as an excellent laboratory for understanding the
properties of quantum chromodynamics (QCD) in the context of a heavy
quark coupling with two light quarks. The ground state,
$\Lambda_{c}^{+}$, was first observed by the MARKII experiment in
1979~\cite{Abrams:1979iu}. In recent years, the BESIII Collaboration
has reported a series of absolute branching fractions (BFs) of
exclusive decays of $\Lambda_{c}^{+}$
~\cite{BESIII:2015ysy,BESIII:2015bjk,BESIII:2016ozn,BESIII:2017fim,BESIII:2016ffj,BESIII:2017rfd,BESIII:2016yrc,BESIII:2018cvs,BESIII:2018cdl,BESIII:2018qyg,BESIII:2020kzc,BESIII:2022bkj,Li:2021iwf}. These
provided significantly improved BF values for the known decay modes,
and some new decay modes were also discovered. The sum of the observed
and predicted BFs of $\Lambda_{c}^{+}$ is approximately
$90\%$~\cite{ParticleDataGroup:2024cfk,Gronau:2018vei}. The
measurements of the BFs of the inclusive decays of $\Lambda_{c}^{+}$
are important for understanding its decay mechanisms and inferring the
extent of the undiscovered decays.

The Cabibbo-favored (CF) decays are the dominant decay modes of
$\Lambda_{c}^{+}$~\cite{Korner:1978ec,Cheng:2021qpd}. According to the
statistical isospin model, the sum of the BFs of both observed and
predicted CF decays of $\Lambda_{c}^{+}$ is
$(83.2\pm4.9)\%$~\cite{Gronau:2018vei}, mainly involving $\Lambda,
\Sigma, \Xi$ and $\overline{K}^{0}$ in the final state.  The inclusive
BFs are useful for calibrating the CF transition amplitude for
$\Lambda_{c}^{+}$ and are particularly essential for the determination
of the $\Lambda_{c}^{+}$
lifetime~\cite{Cheng:2021qpd,Gratrex:2022xpm}.

In 2014, the BESIII experiment accumulated $\ee$ collision data at the
center-of-mass energy $\sqrt{s}=4599.53$ MeV corresponding to an
integrated luminosity of $(586.9\pm0.1\pm3.9)$ pb$^{-1}$, which
initiated many studies on the inclusive decay of
$\Lambda_{c}^{+}$~\cite{BESIII:2018ciw,BESIII:2018mug,BESIII:2020cpu}. In
2020, the BF of $\LctoKsX$ was measured for the first time, giving the result
$(9.9\pm0.6\pm0.4)\%$~\cite{BESIII:2020cpu}, where $X$ means all
possible final state particles. The statistical isospin model
estimates the total BF of exclusive $\Lambda^+_c$ decays containing
$\overline{K}^{0}/K^{0}$ to be $(21.7\pm0.8)\%$, as presented in
Table~\ref{tab:BFs}. However, the summed BF of all observed exclusive
$\Lambda^+_c$ decays containing $\overline{K}^{0}/K^{0}$ only accounts
for a total of $(15.8\pm0.6)\%$. The determination of the absolute BF
of $\LctoKzX$ is an important input for the search for unmeasured
decay modes of $\Lambda^+_c$ and for testing the BFs predicted by the
statistical isospin model.

\begin{table}[h]
\begin{center}
\caption{Observed and predicted BFs of CF exclusive $\Lambda^+_c$
decays containing
$\overline{K}^{0}/K^{0}$~\cite{ParticleDataGroup:2024cfk,Gronau:2018vei},
where the observed BFs are quoted from the Particle Data Group
(PDG)~\cite{ParticleDataGroup:2024cfk} and the predicted BFs are from
Ref.~\cite{Gronau:2018vei} by the statistical isospin model. The BFs
of the $\overline{K}^{0}/K^{0}$ decay modes are obtained by doubling
the corresponding $K_{S}^{0}$ decay modes. At present, there is no evidence
for the possible difference between $K_{S}^{0}$ and $K_{L}^{0}$ production 
in charm baryon decay and this has been neglected in this paper.}\
		\scalebox{0.9}{
		\begin{tabular}{cccc}
			\hline
			Mode & Value & Mode & Value \\
			\hline
			\multicolumn{2}{c}{Observed BF} & \multicolumn{2}{c}{Predicted BF} \\
			\hline
			$p\overline{K}^{0}$ & $(3.18\pm0.14)\%$ & $n\overline{K}^{0}\pi^{+}\pi^{0}$ & $(3.07\pm0.16)\%$ \\
			$p\overline{K}^{0}\pi^{0}$ & $(3.92\pm0.24)\%$ & $p\overline{K}^{0}\pi^{0}\pi^{0}$ & $(1.36\pm0.07)\%$ \\
			$p\overline{K}^{0}\pi^{+}\pi^{-}$ & $(3.18\pm0.22)\%$ & $n\overline{K}^{0}\pi^{+}\pi^{+}\pi^{-}$ & $(0.14\pm0.09)\%$ \\
			$n\overline{K}^{0}\pi^{+}$ & $(3.64\pm0.50)\%$ & $p\overline{K}^{0}\pi^{+}\pi^{-}\pi^{0}$ & $(0.22\pm0.14)\%$ \\
			$p\overline{K}^{0}\eta$ & $(0.88\pm0.06)\%$ & $n\overline{K}^{0}\pi^{+}\pi^{0}\pi^{0}$ & $(0.10\pm0.06)\%$ \\
			$\Lambda\overline{K}^{0}K^{+}$ & $(0.56\pm0.11)\%$ & $p\overline{K}^{0}\pi^{0}\pi^{0}\pi^{0}$ & $(0.03\pm0.02)\%$ \\
			$\Sigma^{+}\phi,\phi\to K_{L}^{0}K_{S}^{0}$ & $(0.13\pm0.02)\%$ & $(\Sigma K)^{+} \overline{K}^{0}$ & $(0.68\pm0.34)\%$ \\
			$\Sigma^{+}K^{*0},K^{*0}\to K^{0}\pi^{0}$ & $(1.16\pm0.33)\times10^{-3}$ & $\Xi^{0}K^{0}\pi^{+}$ & $(0.62\pm0.06)\%$ \\
			$p\overline{K}^{0}K^{0}$ & $(9.40\pm0.72)\times10^{-4}$ &  &  \\
			$\Sigma^{+}K^{0}$ & $(9.40\pm2.80)\times10^{-4}$ &  &  \\
			$p\phi,\phi\to K_{L}^{0}K_{S}^{0}$ & $(3.59\pm0.47)\times10^{-4}$ & & \\
			\hline
			Sum & $(15.8\pm0.6)\%$ & & $(6.2\pm0.4)\%$ \\
			\hline
			Total & \multicolumn{3}{c}{$(22.0\pm0.7)\%$} \\
			\hline
		\end{tabular}
	}
		\label{tab:BFs}
	\end{center}
\end{table}

\begin{table}[hpbt]
\begin{center}
\caption{Center-of-mass energies and corresponding luminosities for data samples used in this work.}\
	
\begin{tabular}{c r@{.}l}
\hline
$\sqrt{s}$ (MeV)	&  \multicolumn{2}{c}{Luminosity (pb$^{-1}$)} \\ \hline
$4599.53\pm0.07\pm0.74$   &   $586.9\pm0$&$1\pm3.9$  \\
$4611.86\pm0.12\pm0.32$   &   $103.8\pm0$&$1\pm0.6$  \\
$4628.00\pm0.06\pm0.32$   &   $521.5\pm0$&$1\pm2.8$  \\
$4640.91\pm0.06\pm0.38$   &   $552.4\pm0$&$1\pm2.9$  \\
$4661.24\pm0.06\pm0.29$   &   $529.6\pm0$&$1\pm2.8$  \\
$4681.92\pm0.08\pm0.29$   &   $1669.3\pm0$&$2\pm8.8$ \\
$4698.82\pm0.10\pm0.39$   &   $536.4\pm0$&$1\pm2.8$  \\
\hline 
\end{tabular}
\label{tab:Luminosity}
\end{center}
\end{table}

This paper presents an improved measurement of the absolute BF of the
inclusive decay $\Lambda^+_c\to K^0_SX$. The measurement is based on
$e^+e^-$ collision data taken with the BESIII detector at
center-of-mass energies between $4599.53$ MeV and $4698.82$ MeV,
corresponding to a total integrated luminosity of $4.5$
fb$^{-1}$~\cite{BESIII:2022xne}. The energy points and corresponding
luminosities are listed in Table~\ref{tab:Luminosity}. These energies
ensure the clean production of the charmed hyperon pairs without
additional hadrons, which facilitates the application of the
double-tag (DT) method, initially introduced by the MARKIII
Collaboration~\cite{MARK-III:1985qfw,MARK-III:1989dea}. Utilizing this
technique, we reconstruct the $\overline{\Lambda}_{c}^{-}$ baryon
through one of the eleven tag modes ($\overline{p}K_{S}^{0}$,
$\overline{p}K^{+}\pi^{-}$, $\overline{p}K_{S}^{0}\pi^{0}$,
$\overline{p}K_{S}^{0}\pi^{-}\pi^{+}$,
$\overline{p}K^{+}\pi^{-}\pi^{0}$, $\overline{\Lambda} \pi^{-}$,
$\overline{\Lambda}\pi^{-}\pi^{0}$, $\overline{\Lambda} \pi^{-}
\pi^{+} \pi^{-}$, $\overline{\Sigma}^{0} \pi^{-}$,
$\overline{\Sigma}^{-} \pi^{0}$ and $\overline{\Sigma}^{-} \pi^{-}
\pi^{+}$), called single-tag (ST)
$\overline{\Lambda}_{c}^{-}$. Subsequently, we identify the
$K_{S}^{0}$ candidate by reconstructing a pair of oppositely charged
tracks that recoil against the ST $\overline{\Lambda}_{c}^{-}$,
thereby forming a DT candidate. Charge conjugation is always implied
throughout this paper.

\section{BESIII Experiment and Monte Carlo Simulation}
\label{sec:detector_dataset}
The BESIII detector~\cite{BESIII:2009fln} records symmetric
$e^{+}e^{-}$ collisions provided by the BEPCII storage
ring~\cite{Yu:2016cof} in the center-of-mass energies range from $2.0$
GeV to $4.95$ GeV, with a peak luminosity of $1.1\times10^{33}$
cm$^{-2}$s$^{-1}$ achieved at a center-of-mass energy of
$\sqrt{s}=3.77$ GeV. The cylindrical core of the BESIII detector
covers $93\%$ of the full solid angle and comprises a helium-based
multilayer drift chamber (MDC), a plastic scintillator time-of-flight
system (TOF), and a CsI(Tl) electromagnetic calorimeter (EMC), which
are all enclosed in a superconducting solenoidal magnet providing a
$1.0$ T magnetic field. The solenoid is supported by an octagonal
flux-return yoke which is segmented into layers and instrumented with
resistive plate counter modules for muon identification. The
charged-particle momentum resolution at $1$ GeV/$c$ is $0.5\%$, and
ionization energy loss d$E/$d$x$ resolution is $6\%$ for electrons
from Bhabha scattering. The EMC measures photon energies with a
resolution of $2.5\%$ ($5\%$) at 1 GeV in the barrel (end cap)
region. The time resolution of TOF barrel region is 68 ps, while that
in the end cap region was 110 ps. The end cap TOF system was updated in
2015 using multigap resistive plate chamber technology, providing a
time resolution of 60
ps~\cite{Li:2017jpg,Guo:2017sjt,Cao:2020ibk}. About $85\%$ of the
$\Lambda_c^{+}\overline{\Lambda}_c^{-}$ pairs are produced in data
taken after this upgrade. More details can be found in
Ref.~\cite{BESIII:2009fln,Yu:2016cof}.

High-statistics Monte Carlo (MC) simulated data samples produced with
{\sc geant4}-based~\cite{GEANT4:2002zbu,Allison:2006ve} software,
which includes the geometric description of the BESIII detector and
the detector response, are used to determine detection efficiencies
and estimate backgrounds.  The $e^{+}e^{-}$ annihilation is simulated
with the {\sc kkmc} generator~\cite{Jadach:2000ir} incorporating the
initial-state radiation (ISR) effects and the beam energy spread.  The
inclusive MC sample includes the
$\Lambda_c^{+}\overline{\Lambda}_c^{-}$ events, $D_{(s)}^{(*)}$
production, ISR return to lower-mass $\psi$ states, and continuum
processes $e^{+}e^{-}\rightarrow q\overline{q}$ ($q=u,d,s$). All the
known decay modes of charmed hadrons and charmonia are modeled with
{\sc evtgen}~\cite{Lange:2001uf, Ping:2008zz} using BFs taken from the
PDG~\cite{ParticleDataGroup:2024cfk}, while the remaining unknown
decays are modeled with {\sc lundcharm}~\cite{Chen:2000tv}.  In
addition, exclusive DT signal MC events, where the
$\overline{\Lambda}_{c}^{-}$ decays into the studied ST channels and
the $\Lambda_{c}^{+}$ decays into $K_{S}^{0} X$, are used to determine
the DT detection efficiencies.  For the MC production of
$\Lambda_c^{+}\overline{\Lambda}_c^{-}$, the Born cross sections are
taken into account~\cite{BESIII:2022xne}, and phase space generated
$\lambdacp$ decays are reweighted according to their observed
distributions.

\section{Event Selection}
%%%%%%     Charged     track     %%%%%%%%%%
Charged tracks detected in the MDC are required to be within a polar
angle $(\theta)$ range of $\lvert \cos\theta \rvert < 0.93$, where
$\theta$ is defined with respect to the $z$ axis, which is the
symmetry axis of the MDC. The distance of the closest approach of the
track to the interaction point (IP) must be less than $10~$cm along
the $z$ axis, and within $1~$cm in the transverse plane of the $z$
axis.  Both the time-of-flight system and the specific ionization
energy loss (d$E/$d$x$) in the MDC are used to determine the
likelihoods $\mathcal{L}$ of different particle type
hypotheses. Tracks are identified as protons when the particle
identification (PID) determines this hypothesis to have the greatest
likelihood ($\mathcal{L}(p)>\mathcal{L}(K)$ and
$\mathcal{L}(p)>\mathcal{L}(\pi)$), while charged kaons and pions are
identified based on comparing the likelihoods for the two hypotheses
($\mathcal{L}(K)>\mathcal{L}(\pi)$ or
$\mathcal{L}(\pi)>\mathcal{L}(K)$). These criteria are not required for
the tracks that are used to reconstruct $K_{S}^0$ or
$\overline{\Lambda}$.

%%%%%%     good      photons     %%%%%%%%%%
Neutral showers are reconstructed in the EMC.  Showers not associated
with any charged track are identified as photon candidates if they
satisfy two additional criteria: (1) an energy deposition in the
EMC of $E_\mathrm{dep} > 25\mev$ in the barrel region corresponding to
the polar angle $\lvert \cos\theta \rvert < 0.8$, while
$E_\mathrm{dep} >50\mev$ in the end-cap region corresponding to $0.86<
\lvert \cos\theta \rvert <0.92$.  (2) the EMC time difference from the
event start time is required to be less than 700 ns to suppress
electronic noise and showers unrelated to the event.  The $\pi^{0}$
candidates are reconstructed from photon pairs with the requirement
that their invariant masses lie within
$115\,\mevcc<M(\gamma\gamma)<150\mevcc$.  To improve the momentum
resolution, a mass-constrained kinematic fit to the $\pi^{0}$ nominal
mass is applied to the photon pairs, and the updated energy and
momentum of the $\pi^0$ are used for the further analysis.

%%%%%%    $K_{S}^{0}$ $\Lambda$$\Sigma , \omega$  Reconstruction   %%%%%%%%%%
Candidates for $K_{S}^{0}$ and $\overline{\Lambda}$ are formed by
combining two oppositely charged tracks into the final states
$\pi^{+}\pi^{-}$ and $\overline{p}\pi^{+}$.  For these two tracks,
their distances of closest approaches to the IP must be within
$\pm$20\,cm along the beam direction.  No distance constraints in the
transverse plane are required.  The charged pion candidate is not
subjected to the PID requirements described above, while PID for the
proton is implemented to improve the quality of the signal.  The two
daughter tracks are constrained to originate from a common decay
vertex, and the $\chi^2$ of the vertex fit is required to be less than
100.  Furthermore, the decay vertex is required to be separated from
the IP by a distance of at least twice the fitted vertex resolution.
The fitted momenta of the $\pi^{+}\pi^{-}$ and $\overline{p}\pi^{+}$
are used in the further analysis.  To select $K_{S}^{0}$ and
$\overline{\Lambda}$ candidates, we impose the requirements
$487\,\mevcc<M(\pi^+\pi^{-})<511\,\mevcc$ and
$1111\,\mevcc<M(\overline{p}\pi^{+})<1121\,\mevcc$, respectively,
which are within about three standard deviations of their mass
resolutions in the MC samples.  The $\overline{\Sigma}^0$ and
$\overline{\Sigma}^-$ candidates are reconstructed with
$\gamma\overline{\Lambda}$ and $\overline{p}\pi^{0}$ with invariant
masses in $1179\,\mevcc<M(\gamma\overline{\Lambda})<1203\mevcc$ and
$1176\,\mevcc<M(\overline{p}\pi^{0})<1200\mevcc$, respectively.

%%%%%%%%%%%%%%%%%%%%%% about cross feed veto  %%%%%%%%%%%%%%%%%%%%
For the tag modes $\Modec$, $\overline{p}K_{S}^{0}\pi^{-}\pi^{+}$, and
$\Modeddd$, possible backgrounds with $\overline{\Lambda}\to
\overline{p}\pip$ are rejected by requiring $M(\overline{p}\pip)$
outside the range $(1110, 1120)\mevcc$.  In addition, in the mode
$\Modec$, candidate events within the range
$1170\,\mevcc<M(\overline{p}\pi^0)<1200\mevcc$ are excluded to
suppress $\overline{\Sigma}^-$ background.  To remove $K_{S}^{0}$
candidates in the modes $\Modedd$, $\Modeccc$, and $\Modeddd$, the
invariant masses of $\pip\pim$ and $\pi^0\pi^0$ pairs are not
allowed to fall in the range $(480, 520)\mevcc$.  The MC simulations
show that peaking backgrounds and cross-feeds among the eleven tag
modes are negligible after performing the above veto procedures.

%%%%%%    $\Lambda_{c}$ Reconstruction   %%%%%%%%%%
The ST $\overline{\Lambda}_{c}^{-}$ yields are identified using the
beam-constrained mass
\begin{eqnarray}
	\begin{aligned}
	 M_{\rm BC}\equiv \sqrt{E_{\rm beam}^2/c^4 - \lvert \vec{p} \rvert^{2}/c^2},
	\end{aligned}
\end{eqnarray}
where $E_{\rm beam}$ is the value of the $e^+$ and $e^-$ beam
energies and $\vec{p}$ is the measured $\overline{\Lambda}_{c}^{-}$
momentum in the center-of-mass system of the $\ee$ collision.  To
improve the signal purity, the energy difference $\Delta{}E \equiv E -
E_{\rm beam}$ for the $\overline{\Lambda}_{c}^{-}$ candidate is
required to comply with a mode-dependent $\Delta{}E$ requirement,
listed in  Table~\ref{tab:deltaE}.
Here, $E$ is the total reconstructed energy of the
$\overline{\Lambda}_{c}^{-}$ candidate.  If more than one candidate
satisfies the above requirements for each ST mode, we select the one
with the minimum $|\Delta{}E|$.  Unbinned maximum likelihood fits are
performed on these $M_{\rm BC}$ distributions to obtain the ST yields,
where the signal shapes are modeled with MC-simulated shapes
convolved with Gaussian functions representing the resolution
difference between data and MC simulation. The background shapes are
described by an ARGUS function~\cite{ARGUS:1990hfq}.

\begin{table}[H]
\begin{center}
\centering
\caption{The $\Delta{}E$ requirement for each tag mode.}\ 
\scalebox{1.0}{
\begin{tabular}{lc}
\hline 
\makebox[0.15\textheight][c]{Tag mode} & \makebox[0.1\textheight][c]{$\Delta{}E$ (MeV)} \\
\hline
$\overline{\Lambda}_{c}^{-} \to \overline{p}K_{S}^{0}$ & $(-21,18)$ \\
$\overline{\Lambda}_{c}^{-} \to \overline{p}K^{+}\pi^{-}$ & $(-29,26)$ \\
$\overline{\Lambda}_{c}^{-} \to \overline{p}K_{S}^{0}\pi^{0}$ & $(-49,34)$ \\
$\overline{\Lambda}_{c}^{-} \to \overline{p}K_{S}^{0}\pi^{-}\pi^{+}$ & $(-34,31)$ \\
$\overline{\Lambda}_{c}^{-} \to \overline{p}K^{+}\pi^{-}\pi^{0}$ & $(-60,41)$ \\
$\overline{\Lambda}_{c}^{-} \to \overline{\Lambda} \pi^{-}$ & $(-23,21)$ \\
$\overline{\Lambda}_{c}^{-} \to \overline{\Lambda}\pi^{-}\pi^{0}$ & $(-50,41)$ \\
$\overline{\Lambda}_{c}^{-} \to \overline{\Lambda} \pi^{-} \pi^{+} \pi^{-}$ & $(-40,36)$ \\
$\overline{\Lambda}_{c}^{-} \to \overline{\Sigma}^{0} \pi^{-}$ & $(-33,31)$ \\
$\overline{\Lambda}_{c}^{-} \to \overline{\Sigma}^{-} \pi^{0}$ & $(-67,32)$ \\
$\overline{\Lambda}_{c}^{-} \to \overline{\Sigma}^{-} \pi^{-} \pi^{+}$ & $(-40,32)$ \\
\hline 
\end{tabular}
}
\label{tab:deltaE}
\end{center}
\end{table}

On the DT side, the $K_{S}^{0}$ candidates are reconstructed from the
remaining tracks on the recoiling side of the tagged
$\overline{\Lambda}_{c}^{-}$. The $K_{S}^{0}$ selection criteria 
are the same as those used in the ST $\overline{\Lambda}_{c}^{-}$
selection. If there is more than one $K_{S}^{0}$ candidate, the one
with the maximum $L/\sigma_{L}$, where $L$ and $\sigma_{L}$ are the
$K^0_S$ decay length from the fit and its associated uncertainty, is
selected for further analysis.

\section{Measurement for Branching Fraction}
\label{sec:bf}

\begin{figure}[]
\begin{center}
\subfigure
{
	\includegraphics[width=0.5\textwidth]{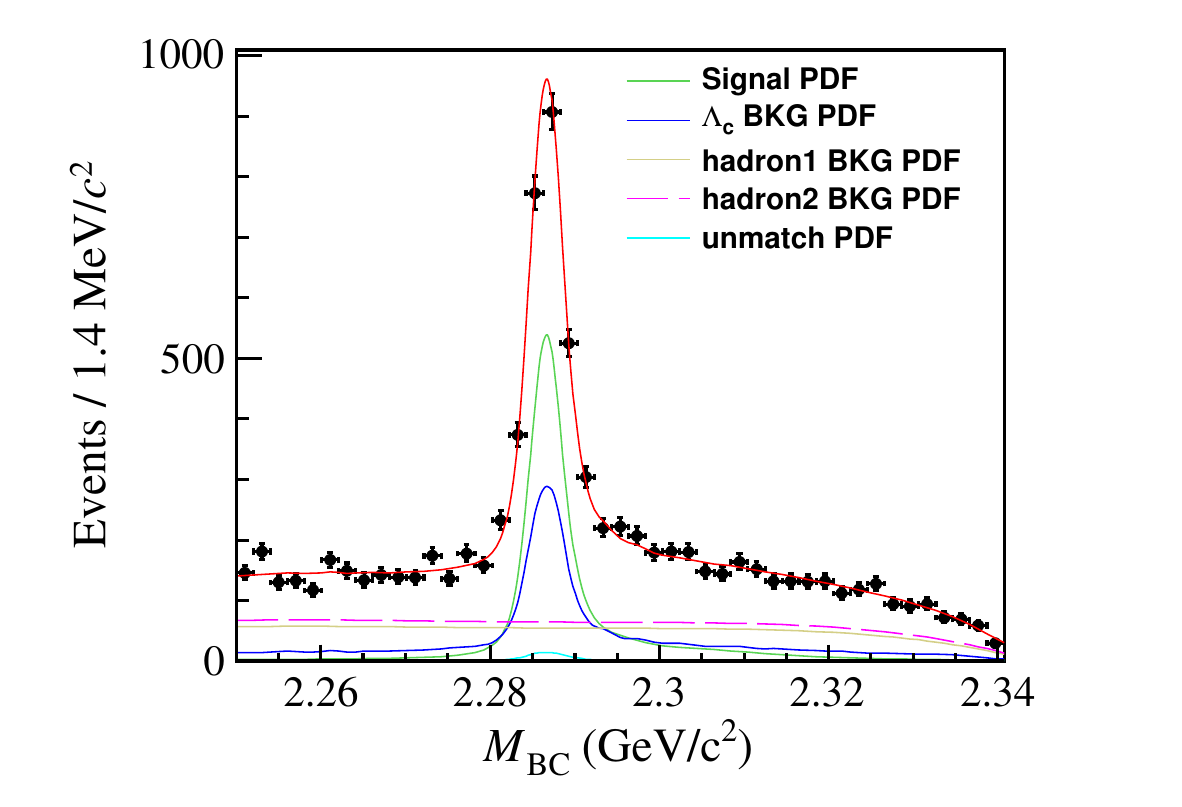}
}\hspace{-10mm}
\subfigure
{
	\includegraphics[width=0.5\textwidth]{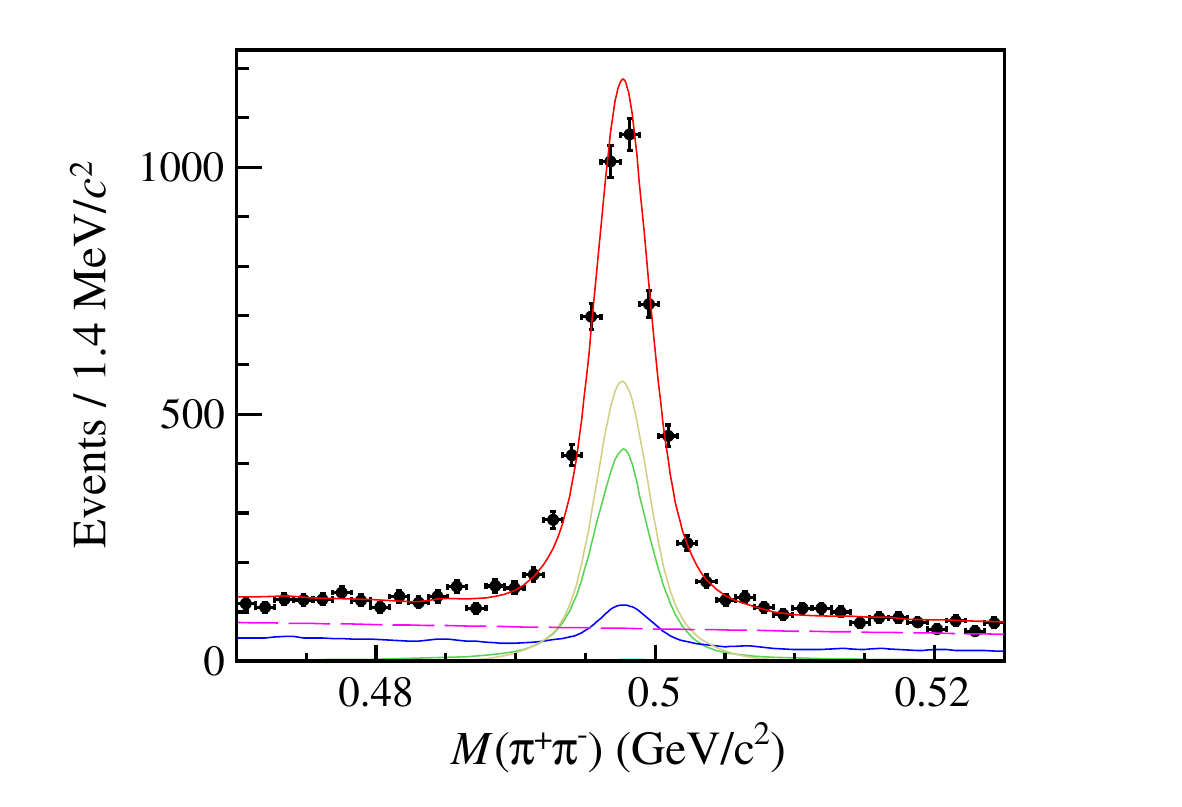}
}
\subfigure
{
	\includegraphics[width=0.5\textwidth]{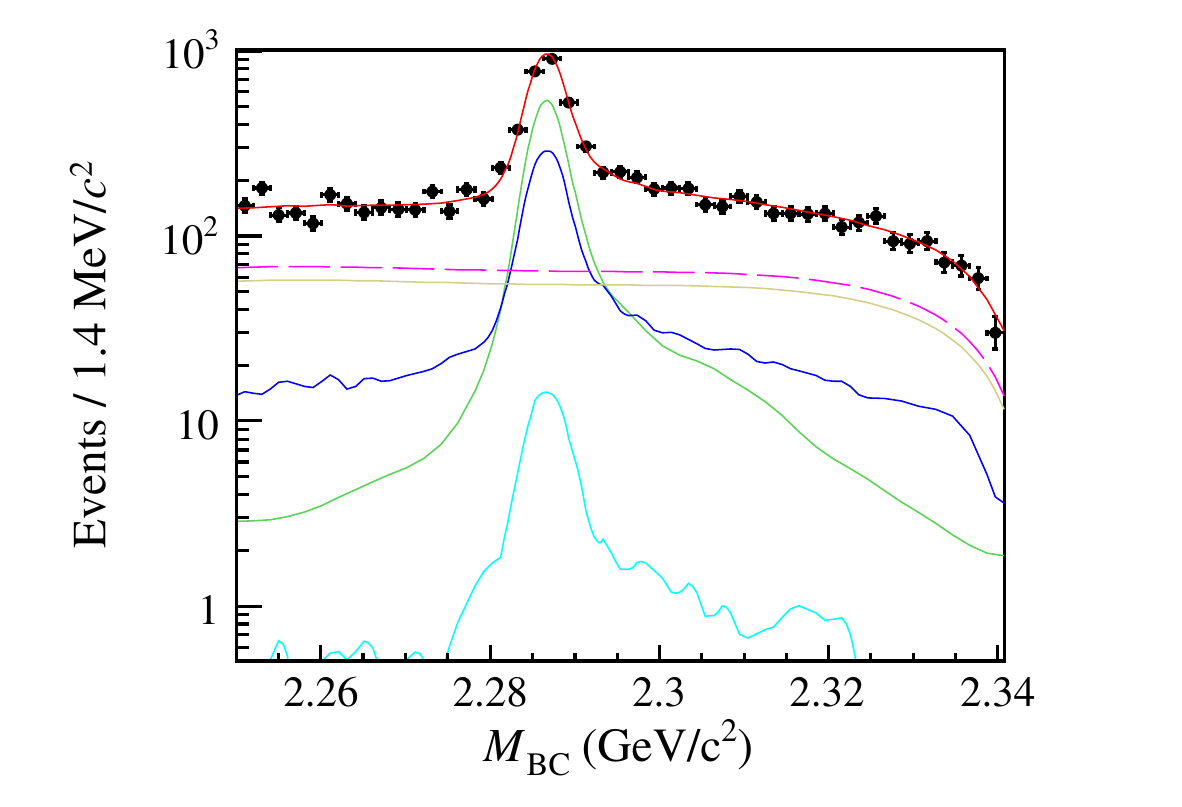}
}\hspace{-10mm}
\subfigure
{
	\includegraphics[width=0.5\textwidth]{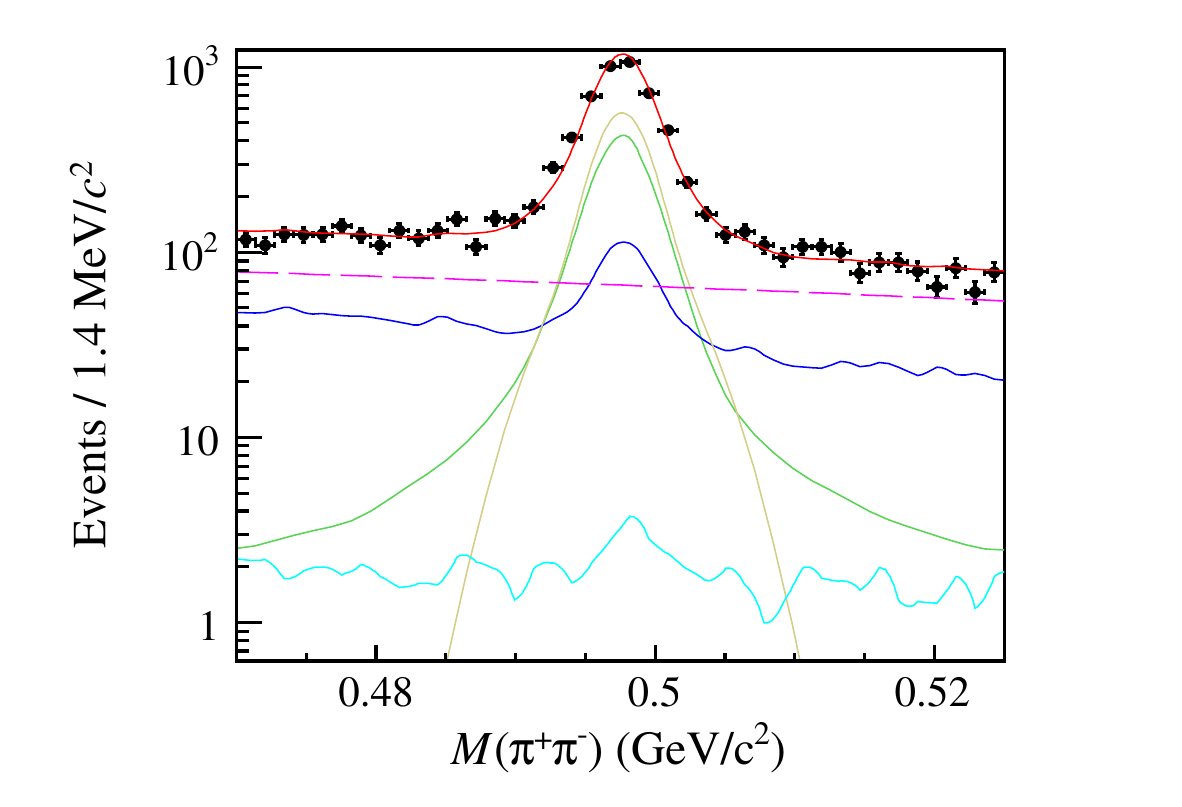}
}
\caption{Projections of the simultaneous 2D fit to the $M_{\rm
BC}$ versus
  $M(\pi^+\pi^-)$ distribution at $\sqrt{s}=4681.92$ MeV. The points with error bars are
  data, the red solid lines are the sums of the fit functions, the green
  solid lines are the $\LctoKsX$ matched signal shapes, the blue solid lines are the $\lambdacp\lambdacm$
  background shapes, the brown solid lines are the hadron1 background
  shapes, the pink dashed lines are the hadron2 background shapes, and the
  cyan solid lines are the unmatched event shapes. The ordinate of the plots in the second row is logarithmic.}  \label{Fig:2dfit}
\end{center}
\end{figure}

A two-dimensional (2D) unbinned maximum likelihood fit to
the distributions of $M_{\rm BC}$ versus the invariant mass of
$\pi^{+}\pi^{-}$, $M(\pi^+\pi^-)$, is performed at the seven energy points
simultaneously to determine the DT signal yields, with the results at $\sqrt{s}=4681.92$ MeV shown in
Fig.~\ref{Fig:2dfit}. 

%The BF of $\LctoKsX$ is a shared parameter for
%the simultaneous fit.

To obtain clean DT signal shapes, a match based on the MC truth
information is performed for signal processes in the inclusive MC
sample. The ratio to evaluate the quality of the match is defined as:
\begin{eqnarray}
	\begin{aligned}
	R_{d\vec{p}} = \dfrac{\lvert \vec{p}_{\rm truth} - \vec{p}_{\rm rec} \rvert}{\lvert \vec{p}_{\rm truth} \rvert},
	\end{aligned}
\end{eqnarray}
where $\vec{p}_{\rm truth}$ and $\vec{p}_{\rm rec}$ are the truth and
reconstructed three-momenta of $\pi^\pm$. For matched events, the
$\pi^{+}\pi^{-}$ from $K_{S}^{0}$ in the DT side are required to
satisfy $R_{d\vec{p}}(\pi^{+}) < 0.5$ and $R_{d\vec{p}}(\pi^{-}) <
0.5$ in signal processes. Otherwise, they are defined as unmatched
events. This match process will bring an efficiency, which 
is embeded in the DT efficiencies.

The DT signal shapes are described by the
MC-simulated shapes of matched events convolved with Gaussian functions, whose
parameters are free, while unmatched events have their own shapes. 
In the fit, the ratios of matched signal events and
unmatched events are fixed based on a study of the signal processes in
the inclusive MC samples.

The background shapes from non-signal $e^{+}e^{-} \to \Lambda_{c}^{+}
\overline{\Lambda}_{c}^{-}$ events are obtained from inclusive MC
samples. The background from $e^{+}e^{-}\rightarrow q\overline{q}$ is
divided into two types: peaking and combinatorial backgrounds in the
$M(\pi^+\pi^-)$ distribution, named hadron1 and hadron2,
respectively. Hadron1 is described by a third-order Chebyshev
polynomial function in the $M_{\rm BC}$ distribution and a double
Gaussian function in the $M(\pi^+\pi^-)$ distribution.  Hadron2 is
described by a third-order Chebyshev polynomial function in the
$M_{\rm BC}$ distribution and a first-order Chebyshev polynomial
function in the $M(\pi^+\pi^-)$ distribution. The parameters of these
functions are obtained by performing a 2D unbinned maximum likelihood
fit to the distribution of $M_{\rm BC}$ versus $M(\pi^+\pi^-)$ of
$e^{+}e^{-}\rightarrow q\overline{q}$~MC samples. The yields of
hadron1 and hadron2 are free parameters.

The BF of $\LctoKsX$ at the individual energy points can be obtained
from the relative yields of DT events to ST events with the correction
of their efficiencies estimated from MC simulations, which is given
by:

\begin{eqnarray}
	\label{cal_BF}
	\mathcal{B_{\alpha}} = \frac{N_{\alpha}^{\rm DT}}{\mathcal{B}_{\rm int}\cdot\sum_{i}(\frac{N_{i}^{\alpha,\rm ST}}{\varepsilon_{i}^{\alpha,\rm ST}}\cdot\varepsilon_{i}^{\alpha,\rm DT})},
\label{eq:bfall}
\end{eqnarray} 
where $N_{\alpha}^{\rm DT}$ is the DT yield, $\mathcal{B}_{\rm int}$
is the BF of $K_{S}^{0} \to
\pi^{+}\pi^{-}$~\cite{ParticleDataGroup:2024cfk}, $N_{i}^{\alpha,\rm
ST}$ is the ST yield and $\varepsilon_{i}^{\alpha,\rm ST}$ is the ST
efficiency, $\varepsilon_{i}^{\alpha,\rm DT}$ is the DT efficiency
estimated by DT signal MC samples,  $i$ represents the index of each
tag mode, and $\alpha$ stands for the label of the energy point.

However, we want to determine the overall BF of $\LctoKsX$, so we
treat the BF as a shared parameter between energy points in the
simultaneous fit and parametrize $N_{\alpha}^{\rm DT}$ in the fit as

\begin{eqnarray} N_{\alpha}^{\rm DT} \to
\mathcal{B}_{\rm int}\cdot\sum_{i}(\frac{N_{i}^{\alpha,\rm
ST}}{\varepsilon_{i}^{\alpha,\rm ST}}\cdot\varepsilon_{i}^{\alpha,\rm
DT})\cdot \mathcal{B}(\LctoKsX),
\end{eqnarray}
and the fit obtains $\mathcal{B}(\LctoKsX)$ directly.

\begin{figure}[H]
	\centering
	\subfigure[]
	{
		\includegraphics[width=0.47\textwidth]{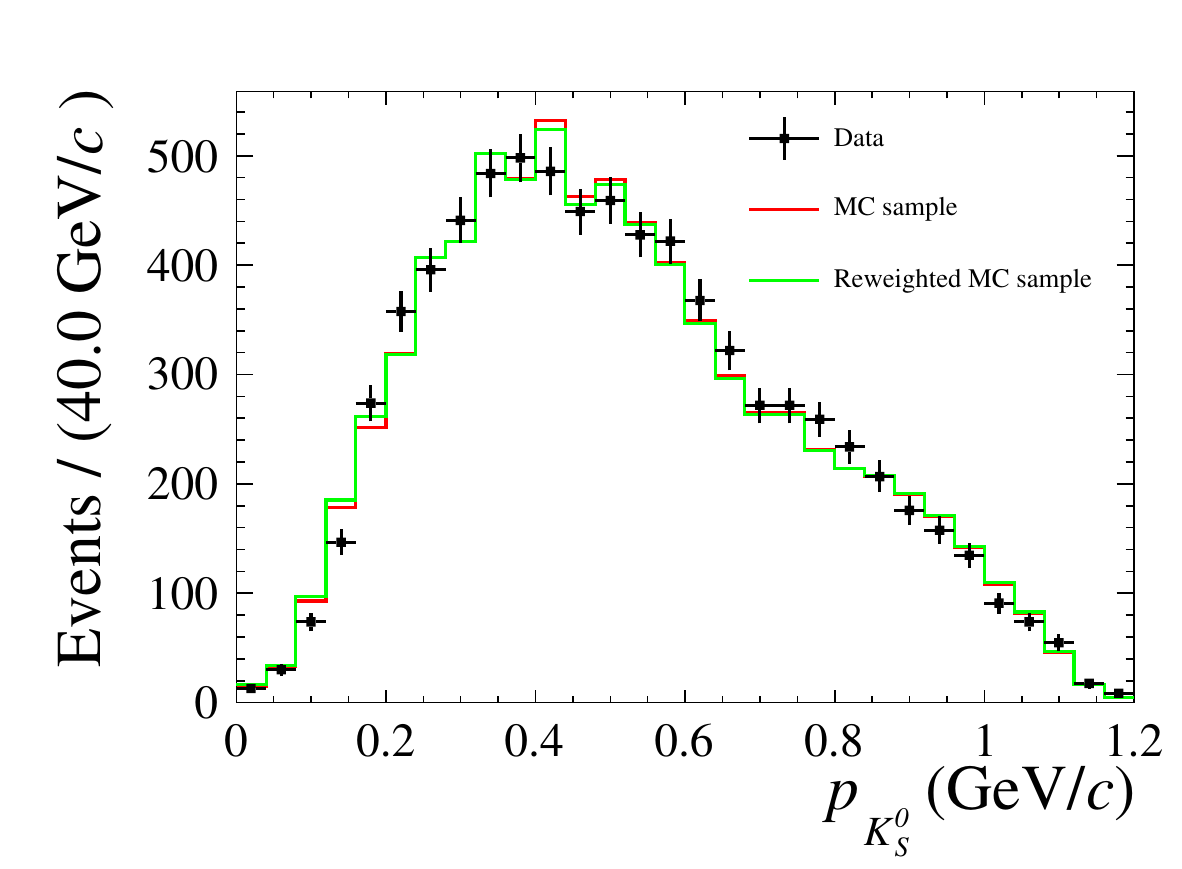}
	}
	\subfigure[]
	{
		\includegraphics[width=0.47\textwidth]{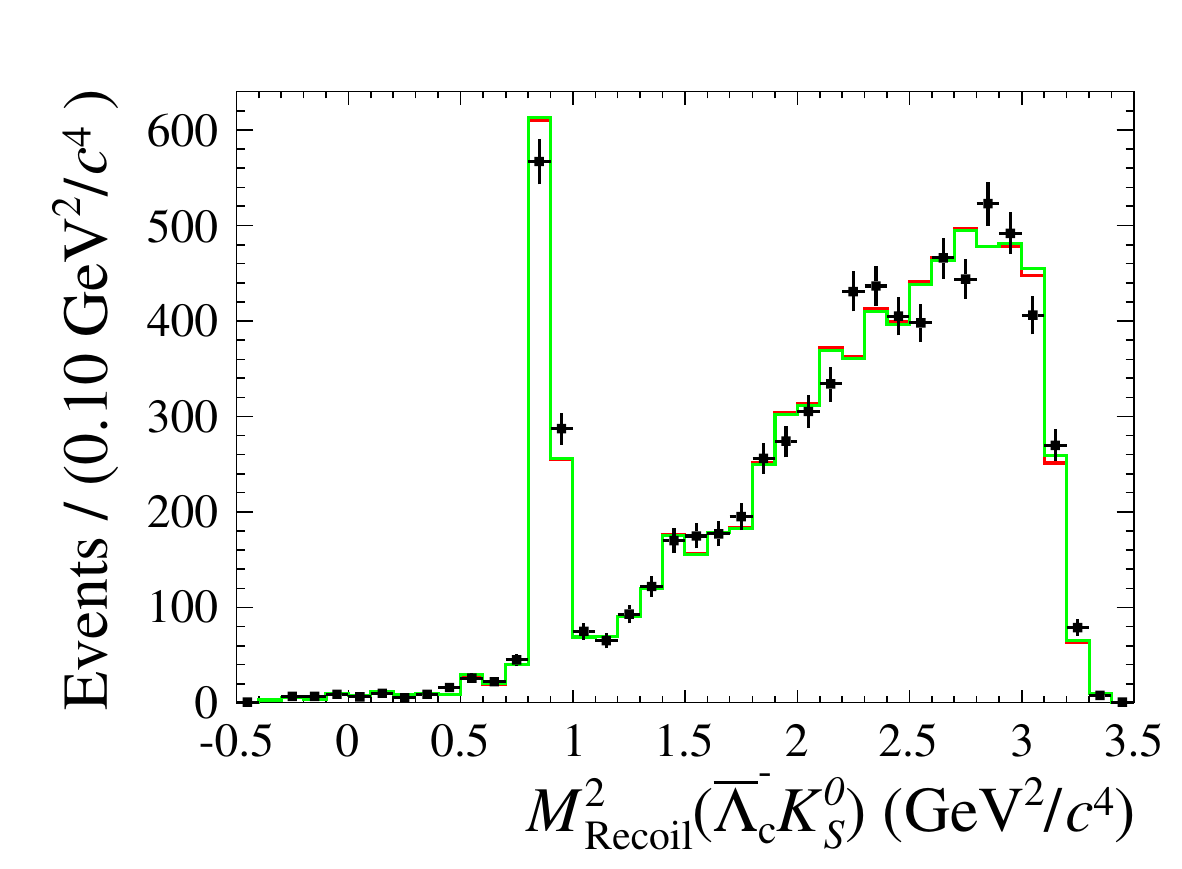}
	}
	
	\caption{The distributions of (a) $p_{K^0_S}$ and (b) $M^2_{\rm Recoil}(\overline\Lambda^{-}_{c}K^0_S)$ of the accepted candidates. The points with error bars are data, the solid red line is MC sample, the solid green line is reweighted MC sample. They are normalized according to the data.}
	\label{fig:comapre}
\end{figure}

To obtain a more accurate DT efficiency, we use the control sample
$J/\psi\to K_{S}^{0}K^{\pm}\pi^{\mp}$ to study the $K_{S}^{0}$
reconstruction efficiency. A factor depending on the $K_{S}^{0}$
momentum distribution is obtained by studying the difference of
$K_{S}^{0}$ reconstruction efficiency between data and MC samples of
$J/\psi\to K_{S}^{0}K^{\pm}\pi^{\mp}$. We use this factor to weight
the DT signal MC samples according to the $K_{S}^{0}$ momentum
distribution in $\LctoKsX$. The relative difference between the corrected
efficiencies and the uncorrected is
$(1.8\pm0.3)\%$. Figure~\ref{fig:comapre} shows the comparisons of the
momentum of the $K^0_S$ candidate and the recoil mass square of the
$\overline\Lambda^-_cK^0_S$ system, $M^2_{\rm
Recoil}(\overline\Lambda^-_cK^0_S)$, where a clear peak of proton can be found.
Good agreement between data and
MC simulation can be seen.

 The determination of ST yields and ST
efficiencies are the same as in Ref.~\cite{BESIII:2022xne}. The DT efficiencies of matched events are listed in
Table~\ref{tab:DT_Eff},
and the fitted DT signal yields
of $\LctoKsX$ at the seven energy points are listed in
Table~\ref{tab:DT_yields}.  
 Finally, the BF of
$\Lambda^+_c\to K^0_SX$ is determined to be $(10.9\pm0.2)\%$, where
the uncertainty is statistical only. To verify this value is stable, we measured the BFs at different energy points combining all tag modes, listed in Table~\ref{tab:BF_ene}. And the BFs at each tag mode combining all energy points samples, listed in Table~\ref{tab:BF_tag}.

\begin{table}[H]
	%\footnotesize
	\begin{center}
		\centering
		\caption{The DT efficiencies of matched events $\varepsilon_{i}^{\alpha,\rm DT}$ for each tag mode at various energy points, where the uncertainties are statistical only.}\
		
		\scalebox{0.65}{
			\begin{tabular}{lccccccc}
				\hline 
				\makebox[0.15\textheight][c]{$\varepsilon_{i}^{\alpha,\rm DT}(\%)$} & \makebox[0.1\textheight][c]{$4599.53$ MeV} & \makebox[0.1\textheight][c]{$4611.86$ MeV} & \makebox[0.1\textheight][c]{$4628.00$ MeV} & \makebox[0.1\textheight][c]{$4640.91$ MeV} &
				\makebox[0.1\textheight][c]{$4661.24$ MeV} & \makebox[0.1\textheight][c]{$4681.92$ MeV} & \makebox[0.1\textheight][c]{$4698.82$ MeV} \\
				\hline
				$\overline{\Lambda}_{c}^{-} \to \overline{p}K_{S}^{0}$ 
				& $32.5\pm0.5$ & $30.0\pm1.2$ & $29.1\pm0.5$ & $28.1\pm0.5$ & $28.5\pm0.5$ & $26.8\pm0.3$ & $25.0\pm0.5$ \\
				$\overline{\Lambda}_{c}^{-} \to \overline{p}K^{+}\pi^{-}$ 
				& $29.1\pm0.2$ & $28.1\pm0.5$ & $26.8\pm0.2$ & $26.8\pm0.2$ & $26.2\pm0.2$ & $25.7\pm0.1$ & $25.2\pm0.2$ \\
				$\overline{\Lambda}_{c}^{-} \to \overline{p}K_{S}^{0}\pi^{0}$
				& $12.5\pm0.3$ & $12.6\pm0.8$ & $11.0\pm0.3$ & $11.1\pm0.3$ & $10.7\pm0.3$ & $10.5\pm0.2$ & $10.6\pm0.3$ \\
				$\overline{\Lambda}_{c}^{-} \to \overline{p}K_{S}^{0}\pi^{-}\pi^{+}$ 
				& $12.2\pm0.3$ & $11.2\pm0.6$ & $10.9\pm0.3$ & $11.0\pm0.3$ & $10.7\pm0.3$ & $10.6\pm0.2$ & $10.1\pm0.3$ \\
				$\overline{\Lambda}_{c}^{-} \to \overline{p}K^{+}\pi^{-}\pi^{0}$ 
				& $11.1\pm0.2$ & $10.3\pm0.4$ & $9.5\pm0.2$ & $9.4\pm0.2$ & $9.4\pm0.2$ & $8.9\pm0.1$ & $9.3\pm0.2$ \\
				$\overline{\Lambda}_{c}^{-} \to \overline{\Lambda} \pi^{-}$
				& $26.6\pm0.6$ & $27.8\pm1.4$ & $23.4\pm0.6$ & $22.7\pm0.5$ & $22.6\pm0.6$ & $22.8\pm0.3$ & $21.6\pm0.6$ \\
				$\overline{\Lambda}_{c}^{-} \to \overline{\Lambda}\pi^{-}\pi^{0}$ 
				& $12.2\pm0.2$ & $10.7\pm0.4$ & $10.3\pm0.2$ & $10.5\pm0.2$ & $10.2\pm0.2$ & $9.7\pm0.1$ & $9.4\pm0.2$ \\
				$\overline{\Lambda}_{c}^{-} \to \overline{\Lambda} \pi^{-} \pi^{+} \pi^{-}$
				& $7.9\pm0.2$ & $7.2\pm0.5$ & $7.3\pm0.2$ & $7.6\pm0.2$ & $7.4\pm0.2$ & $7.2\pm0.1$ & $7.3\pm0.2$ \\
				$\overline{\Lambda}_{c}^{-} \to \overline{\Sigma}^{0} \pi^{-}$ 
				& $16.2\pm0.5$ & $15.7\pm1.1$ & $15.0\pm0.5$ & $15.1\pm0.5$ & $13.6\pm0.5$ & $14.8\pm0.3$ & $13.4\pm0.5$ \\
				$\overline{\Lambda}_{c}^{-} \to \overline{\Sigma}^{-} \pi^{0}$ 
				& $13.5\pm0.5$ & $12.2\pm1.1$ & $13.3\pm0.5$ & $13.8\pm0.5$ & $12.5\pm0.5$ & $12.1\pm0.3$ & $11.0\pm0.5$\\
				$\overline{\Lambda}_{c}^{-} \to \overline{\Sigma}^{-} \pi^{-} \pi^{+}$ 
				& $14.2\pm0.3$ & $14.4\pm0.6$ & $13.4\pm0.3$ & $13.3\pm0.3$ & $12.4\pm0.3$ & $12.8\pm0.2$ & $12.4\pm0.3$\\
				\hline 
			\end{tabular}
		}
		\label{tab:DT_Eff}
	\end{center}
\end{table}

\begin{table}[H]
\begin{center}
\caption{The fitted DT yields $N_{\alpha}^{\rm DT}$ at various energy points.}\
		
\begin{tabular}{c c}
\hline
$\sqrt s$~(MeV) & $N_{\alpha}^{\rm DT}$ \\
\hline
4599.53 & $669\pm38$  \\
4611.86 & $132\pm17$  \\
4628.00 & $602\pm36$  \\
4640.91 & $605\pm36$  \\
4661.24 & $691\pm39$  \\
4681.92 & $1733\pm62$ \\
4698.82 & $507\pm33$  \\
\hline
Total & $4939\pm104$  \\
\hline 
\end{tabular}
\label{tab:DT_yields}
\end{center}
\end{table}

\begin{table}[H]
	\begin{center}
		\caption{The BFs at various energy points.}\
		
		\begin{tabular}{c c}
			\hline
			$\sqrt s$~(MeV) & $\mathcal{B}(\LctoKsX)(\%)$ \\
			\hline
			4599.53 & $10.0\pm0.6$  \\
			4611.86 & $11.3\pm1.4$  \\
			4628.00 & $11.0\pm0.7$  \\
			4640.91 & $11.0\pm0.6$  \\
			4661.24 & $11.2\pm0.7$  \\
			4681.92 & $10.9\pm0.4$ \\
			4698.82 & $11.1\pm0.7$  \\
			\hline
			Total & $10.9\pm0.2$  \\
			\hline 
		\end{tabular}
		\label{tab:BF_ene}
	\end{center}
\end{table}

\begin{table}[H]
	\begin{center}
		\caption{The BFs at various tag modes.}\
		
		\begin{tabular}{c c}
			\hline
			Tag mode & $\mathcal{B}(\LctoKsX)(\%)$ \\
			\hline
			$\overline{\Lambda}_{c}^{-} \to \overline{p}K_{S}^{0}$ & $11.1\pm0.8$ \\
			$\overline{\Lambda}_{c}^{-} \to \overline{p}K^{+}\pi^{-}$ & $10.9\pm0.4$ \\
			$\overline{\Lambda}_{c}^{-} \to \overline{p}K_{S}^{0}\pi^{0}$ & $11.3\pm1.1$ \\
			$\overline{\Lambda}_{c}^{-} \to \overline{p}K_{S}^{0}\pi^{-}\pi^{+}$ & $10.9\pm1.2$ \\
			$\overline{\Lambda}_{c}^{-} \to \overline{p}K^{+}\pi^{-}\pi^{0}$ & $10.7\pm0.7$ \\
			$\overline{\Lambda}_{c}^{-} \to \overline{\Lambda} \pi^{-}$ & $10.8\pm1.0$ \\
			$\overline{\Lambda}_{c}^{-} \to \overline{\Lambda}\pi^{-}\pi^{0}$ & $10.8\pm0.7$ \\
			$\overline{\Lambda}_{c}^{-} \to \overline{\Lambda} \pi^{-} \pi^{+} \pi^{-}$ & $11.2\pm1.0$ \\
			$\overline{\Lambda}_{c}^{-} \to \overline{\Sigma}^{0} \pi^{-}$ & $11.3\pm1.3$ \\
			$\overline{\Lambda}_{c}^{-} \to \overline{\Sigma}^{-} \pi^{0}$ & $10.9\pm1.6$ \\
			$\overline{\Lambda}_{c}^{-} \to \overline{\Sigma}^{-} \pi^{-} \pi^{+}$ & $10.8\pm0.8$ \\
			\hline 
			Total & $10.9\pm0.2$ \\
			\hline
		\end{tabular}
		\label{tab:BF_tag}
	\end{center}
\end{table}

\section{Systematic Uncertainty}
The systematic uncertainties arising from the ST side  mostly
cancel in the BF measurement according to Eq.~\ref{eq:bfall}. The
systematic uncertainties from various sources are
summarized in Table~\ref{tab:sys_err} and discussed below. 
 
\begin{itemize}
\item $K_{S}^{0}$ reconstruction.  

As described in Section~\ref{sec:bf}, $J/\psi\to
K_{S}^{0}K^{\pm}\pi^{\mp}$ decay is used as a control sample to
determine the correction to the DT efficiencies of $(1.8\pm0.3)\%$.
The systematic uncertainty in the $K_{S}^{0}$ reconstruction is
$0.3\%$.
	\item 2D fitting.
	The uncertainty arising from the simultaneous 2D fitting is estimated by convolving the signal MC shape with a double Gaussian function to vary the signal shape, while the background from $e^{+}e^{-}\rightarrow q\overline{q}$ is described using the shape extracted from $e^{+}e^{-}\rightarrow q\overline{q}$ MC samples. The total uncertainty from the 2D fitting is $0.9\%$.
	\item Intermediate BF.
	The BF of $K_{S}^{0} \to \pi^{+}\pi^{-}$, $\mathcal{B}_{\rm int} = (69.20\pm0.05)\%$~\cite{ParticleDataGroup:2024cfk}, gives an uncertainty of $0.1\%$.
	\item MC statistics.
	The statistical uncertainties of ST efficiencies and DT efficiencies are propagated to the BFs of signal channel according to Eq.~\ref{cal_BF}, and contribute with an uncertainty of $0.3\%$.
	\item ST $\lambdacm$ yield.
	The systematic uncertainty in the total ST yield arises from the background fluctuation together with a component coming from the fit to the $M_{\rm BC}$ distribution. It is studied by varying the parameters of signal shape, fitting range and endpoint of the ARGUS function, and then redoing the fit process. This gives an uncertainty of $0.4\%$.
\item  $R_{d\vec{p}}$ requirement.
The uncertainty arising from the  $R_{d\vec{p}}$ is estimated by
obtaining the BF without this requirement, leading to an uncertainty
of 0.1\%.  
\end{itemize}
The total systematic uncertainty is $1.1\%$ by summing all the uncertainties in quadrature, assuming no correlation exists between these sources.
\begin{table}[H]
\begin{center}
\caption{Relative systematic uncertainties for the measured BF.}\
	
\begin{tabular}{lc}
\hline
Source                                 & $\LctoKsX~(\%)$ \\
\hline
$K_{S}^{0}$ reconstruction               & 0.3              \\
2D fitting                             & 0.9               \\
Intermediate BF                        & 0.1               \\
MC statistics                          & 0.3               \\
ST $\lambdacm$ yield & 0.4               \\
$R_{d\vec{p}}$ requirement   &  0.1  \\ 
\hline
Total                                  & 1.1 \\
\hline 
\end{tabular}
\label{tab:sys_err}
\end{center}
\end{table}

\section{Summary}
The absolute BF of $\LctoKsX$ is measured based on about 4.5 fb$^{-1}$
of $e^+e^-$ collision data collected at center-of-mass energies
ranging from 4599.56 MeV to 4698.82 MeV with the BESIII detector. The
result is $\mathcal{B}(\LctoKsX)=(10.9 \pm0.2 \pm0.1)\%$, where the
first uncertainty is statistical and the second is systematic. The
precision of the measured BF is improved by a factor of three compared to
the previous BESIII measurement
$\mathcal{B}(\LctoKsX)=(9.9\pm0.6\pm0.4)\%$~\cite{BESIII:2020cpu}. Compared
to the summed BF of the observed exclusive $\Lambda^+_c$ decays,
$(7.9\pm0.3)\%$, our result indicates that the combined BF of the undiscovered
decay channels of $\Lambda_{c}^{+}$ that include a $K_{S}^{0}$ meson
in the final state is $(3.0\pm0.4)\%$.  The BF of the inclusive decay
$\LctoKzX$ is calculated to be $(21.8 \pm 0.4 \pm 0.2)\%$. 
This result is consistent with the prediction
from the statistical isospin model, $\mathcal{B}(\LctoKzX) = (22.0 \pm
0.7)\%$. The total predicted BF of unseen decays by the statistical isospin model of
the inclusive decay $\LctoKzX$ is (6.2$\pm$0.4)\% as listed in
Table~\ref{tab:BFs}. This is also consistent with the BF of the
undiscovered decay channels of $\Lambda_{c}^{+}$ containing
$K_{S}^{0}$ in the final state, estimated from our measured result. It
can be seen from Table~\ref{tab:BFs} that the predicted undiscovered
decay channels are dominated by the decays which contain neutrons in
the final state. Due to the challenge of neutron reconstruction, only
the BF of $\Lambda_{c}^{+} \to n K_{S}^{0} \pi^{+}$ has been reported
by BESIII~\cite{BESIII:2016yrc,BESIII:2023pia}. By developing neutron 
reconstruction techniques~\cite{BESIII:2024mgg}, the $\Lambda_{c}^{+} \to
n\overline{K}^{0}\pi^{+}\pi^{0}$ decay, which accounts for almost half
of the unobserved decays of $\LctoKzX$, could be measured. For
$\Lambda_{c}^{+} \to \Xi^{0}K^{0}\pi^{+}$, the theory based on SU(3)
flavor symmetry predicts its BF to be
$(8.7\pm1.7)\%$~\cite{Geng:2019awr}, which differs significantly from
the estimation by the statistical isospin model, as listed in
Table~\ref{tab:BFs}. Our result disfavors the predictions of SU(3)
flavor symmetry indirectly and shows that a direct measurement of the
unobserved $K_{S}^{0}$-involved decays of $\Lambda_{c}^{+}$ is
important for further testing the statistical isospin model.

\acknowledgments
%% Saved at => 2024-11-21
%\textbf{Acknowledgement}
The BESIII Collaboration thanks the staff of BEPCII and the IHEP computing center for their strong support. This work is supported in part by National Key R\&D Program of China under Contracts Nos. 2023YFA1606000, 2023YFA1609400; National Natural Science Foundation of China (NSFC) under Contracts Nos. 12105127, 12305105, 12422504, 11635010, 11735014, 11935015, 11935016, 11935018, 12025502, 12035009, 12035013, 12061131003, 12192260, 12192261, 12192262, 12192263, 12192264, 12192265, 12221005, 12225509, 12235017, 12361141819, 12375092; the Chinese Academy of Sciences (CAS) Large-Scale Scientific Facility Program; the CAS Center for Excellence in Particle Physics (CCEPP); Joint Large-Scale Scientific Facility Funds of the NSFC and CAS under Contract No. U1832207; CAS under Contract No. YSBR-101; 100 Talents Program of CAS; Fundamental Research Funds for the Central Universities, Lanzhou University; The Institute of Nuclear and Particle Physics (INPAC) and Shanghai Key Laboratory for Particle Physics and Cosmology; Agencia Nacional de Investigación y Desarrollo de Chile (ANID), Chile under Contract No. ANID PIA/APOYO AFB230003; German Research Foundation DFG under Contract No. FOR5327; Istituto Nazionale di Fisica Nucleare, Italy; Knut and Alice Wallenberg Foundation under Contracts Nos. 2021.0174, 2021.0299; Ministry of Development of Turkey under Contract No. DPT2006K-120470; National Research Foundation of Korea under Contract No. NRF-2022R1A2C1092335; National Science and Technology fund of Mongolia; National Science Research and Innovation Fund (NSRF) via the Program Management Unit for Human Resources \& Institutional Development, Research and Innovation of Thailand under Contract No. B50G670107; Polish National Science Centre under Contract No. 2019/35/O/ST2/02907; Swedish Research Council under Contract No. 2019.04595; The Swedish Foundation for International Cooperation in Research and Higher Education under Contract No. CH2018-7756; U. S. Department of Energy under Contract No. DE-FG02-05ER41374

%\vspace{-0.4cm}

\bibliographystyle{JHEP}
\bibliography{references_v13}

\clearpage
\appendix
%\author{The BESIII Collaboration}
%% Saved at => 2024-11-21
M.~Ablikim$^{1}$, M.~N.~Achasov$^{4,c}$, P.~Adlarson$^{76}$, X.~C.~Ai$^{81}$, R.~Aliberti$^{35}$, A.~Amoroso$^{75A,75C}$, Q.~An$^{72,58,a}$, Y.~Bai$^{57}$, O.~Bakina$^{36}$, Y.~Ban$^{46,h}$
% ...
\\
\vspace{0.2cm}
(BESIII Collaboration)\\
\vspace{0.2cm} {\it
$^{1}$ Institute of High Energy Physics, Beijing 100049, People's Republic of China\\
% ...
$^{4}$ Budker Institute of Nuclear Physics SB RAS (BINP), Novosibirsk 630090, Russia\\
% ...
$^{35}$ Johannes Gutenberg University of Mainz, Johann-Joachim-Becher-Weg 45, D-55099 Mainz, Germany\\
$^{36}$ Joint Institute for Nuclear Research, 141980 Dubna, Moscow region, Russia\\
% ...
$^{46}$ Peking University, Beijing 100871, People's Republic of China\\
% ...
$^{57}$ Southeast University, Nanjing 211100, People's Republic of China\\
$^{58}$ State Key Laboratory of Particle Detection and Electronics, Beijing 100049, Hefei 230026, People's Republic of China\\
% ...
$^{72}$ University of Science and Technology of China, Hefei 230026, People's Republic of China\\
% ...
$^{75}$ University of Turin and INFN, (A)University of Turin, I-10125, Turin, Italy; (B)University of Eastern Piedmont, I-15121, Alessandria, Italy; (C)INFN, I-10125, Turin, Italy\\
$^{76}$ Uppsala University, Box 516, SE-75120 Uppsala, Sweden\\
% ...
$^{81}$ Zhengzhou University, Zhengzhou 450001, People's Republic of China\\

\vspace{0.2cm}
$^{a}$ Deceased\\
% ...
$^{c}$ Also at the Novosibirsk State University, Novosibirsk, 630090, Russia\\
% ...
$^{h}$ Also at State Key Laboratory of Nuclear Physics and Technology, Peking University, Beijing 100871, People's Republic of China\\
% ...

}
%% ends here %%

\end{document}